\documentclass[traditabstract]{aa}
\usepackage{graphicx}
\usepackage{txfonts}
\usepackage{longtable}
\usepackage[authoryear]{natbib}
\usepackage{color}
\bibpunct{(}{)}{,}{a}{}{,}
\def\ltsima{$\, \buildrel < \over \sim \,$}
\def\simlt{\lower.5ex\hbox{\ltsima}}
\def\gtsima{$\, \buildrel > \over \sim \,$}
\def\simgt{\lower.5ex\hbox{\gtsima}}

\newcommand{\HI}{\ion{H}{i}}
\newcommand{\HII}{\ion{H}{ii}}

\begin{document}
   \title{The strange case of the peculiar spiral galaxy NGC~5474}
\subtitle{New pieces of a galactic puzzle}

   \author{ M. Bellazzini\inst{1}, F. Annibali\inst{1}, M. Tosi\inst{1}, A. Mucciarelli\inst{2}, M. Cignoni\inst{3}, G. Beccari\inst{4}, C. Nipoti\inst{2} \and R. Pascale\inst{2,1}}
         
      \offprints{M. Bellazzini}

   \institute{INAF - Osservatorio di Astrofisica e Scienza dello Spazio di Bologna,
              Via Gobetti 93/3, 40129 Bologna, Italy
             \email{michele.bellazzini@inaf.it} 
             \and
             Dipartimento di Fisica e Astronomia, Universit\`a degli Studi di Bologna, Via Gobetti93/2, 40129 Bologna, Italy
             \and
             Dipartimento di Fisica, Universit\`a of Pisa, Largo B. Pontecorvo 3, I-56127 Pisa, Italy 0000-0001-6291-6813
              \and
             European Southern Observatory, Karl-Schwarzschild-Strasse 2, D-85748 Garching bei M\"unchen, Germany)}

     \authorrunning{M. Bellazzini et al.}
   \titlerunning{The strange case of NGC~5474 }

   \date{Accepted for publication on A\&A }

\abstract{We present the first analysis of the stellar content of the structures and substructures identified in the peculiar star-forming galaxy NGC~5474, based on Hubble Space Telescope resolved photometry from the LEGUS survey. NGC~5474 is a satellite of the giant spiral M~101, and is known to have a prominent bulge that is significantly off-set from the kinematic center of the underlying \HI ~and stellar disc. The youngest stars (age$\le 100$~Myr) trace a flocculent spiral pattern extending out to $\ga 8$~kpc from the center of the galaxy. On the other hand intermediate-age (age$\ga 500$~Myr) and old (age $\ge 2$~Gyr) stars dominate the off-centred bulge and a large substructure residing in the South Western part of the disc and not correlated with the spiral arms (SW over-density). The old age of the stars in the SW over-density suggests that this may be another signature of the dynamical interaction/s that have shaped this anomalous galaxy. We suggest that a fly by with M~101, generally invoked as the origin of the anomalies, may not be sufficient to explain all the observations. A more local and more recent interaction may help to put all the pieces of this galactic puzzle together.}


\maketitle
%
%________________________________________________________________

\section{Introduction}
\label{intro}

NGC~5474\footnote{Also known as UGC~9013 and VV344b, where VV344a is M101.} is a satellite of M~101 (the Pinwheel galaxy) classified as SAcd pec. It is located at 6.98~Mpc from us \citep{cflow2}, in the M~101 group \citep{cflow3}, at 
0.74$\degr$ from the center of M~101, corresponding to $\simeq 90$~kpc in projection, and 0.80$\degr$ from NGC~5477, another member of the group. Its absolute integrated visual magnitude, $M_V\simeq -18.5$, is $\simeq 1.5$ mag brighter than the conventional limit for dwarf galaxies, adopted, e.g., by \citet{tht}. The total visual luminosity of NGC~5474 is about 1.5 times that of the Large Magellanic Cloud \citep{mc12}.  

\begin{figure}
   \centering
   \includegraphics[width=\columnwidth]{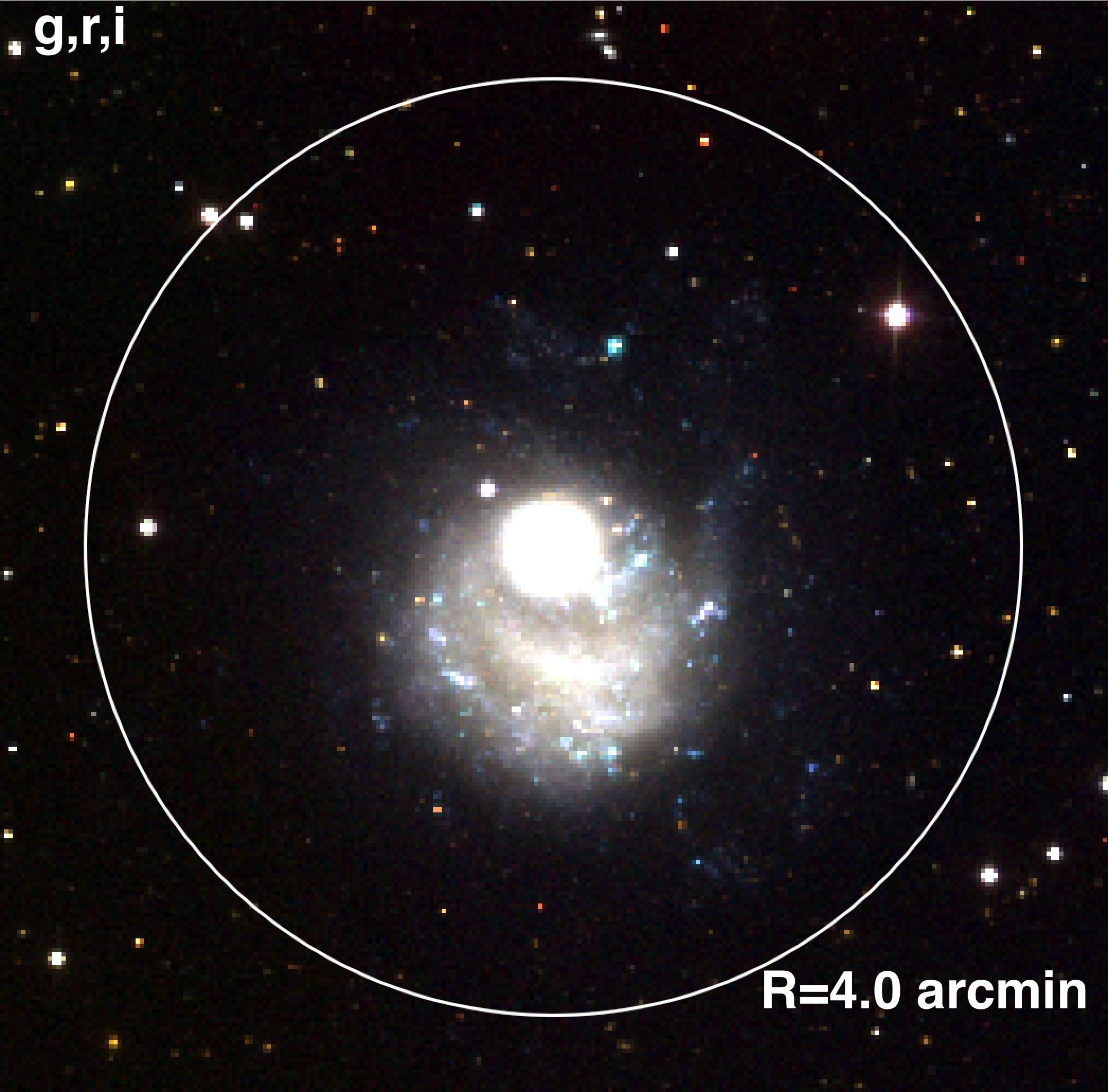}
     \caption{RGB color image of NGC~5474 obtained with SDSS images in g, r, and i band, for the blue, green and red channels, respectively. North is up and East is to the left. The white circle is centred on the center of the bulge.
        \label{foto1}}
    \end{figure}

The most obvious reason for being classified as {\em peculiar} is readily visible in the image shown in Fig.~\ref{foto1}, where a seemingly circular compact bulge (hereafter ``the bulge'', for brevity) is clearly seen to lie at the northern edge of what appears as a bright, nearly face-on, stellar disc. Moreover any additional piece of observational evidence seems to add more complexity to the overall picture of this system \citep[see, e.g,][for discussion and references]{rownd,gonza,korn98,korn00,epi,mihos_OP}. 

%------------------------FIG -----------------------------------

%%%%%%%%%%%%%%%%%%%%%%%%%%%%%%%%%%%%%%%%%%%%%%%%%%%%%%%%%%%%%%%%%

\begin{figure*}
   \centering
   \includegraphics[width=\textwidth]{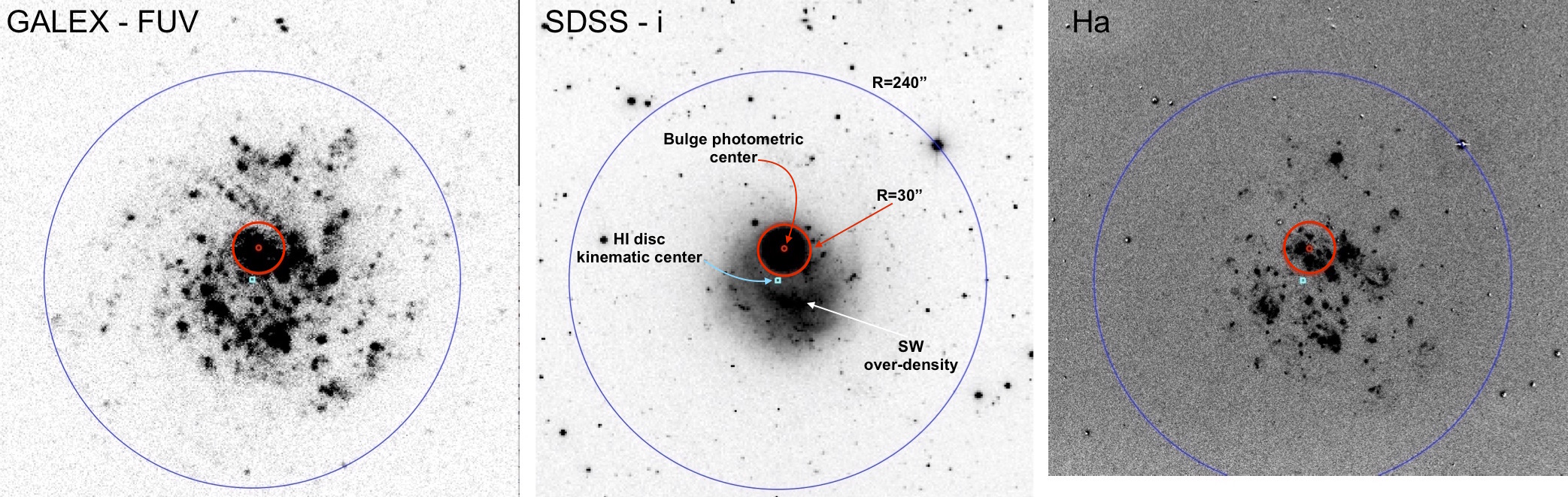}
     \caption{Images of NGC~5474 at different wavelengths. Left: GALEX-FUV, center: SDSS-i, right: continuum-subtracted H$\alpha$ image \citep[retrieved from NED, where it is attributed to][]{dale}. All the images have the same scale and are aligned. 
     The small red circle is the photometric center of the bulge, the wide red circle has radius=$30.0\arcsec$ and encloses the bulge, the small cyan square is the kinematic center of the \HI ~disc \citep[according to][]{rownd}, and the wide blue circle is centred on this point and has a radius=$240.0\arcsec$. North is up and East is to the left.
        \label{foto2}}
    \end{figure*}

The three panels of Fig.~\ref{foto2}, showing images of the galaxy at different wavelengths, provide the basis for a more thorough illustration of most of the anomalies affecting NGC~5474. In the central panel we have marked and labelled some remarkable features, for reference. In particular, the small red circle marks the optical center of the bulge, that is easy to locate (see, e.g., the intensity contours in Fig.~\ref{hima}, below) and coincides with the position of a nuclear star cluster (see Sect.~\ref{sec_nucleus}).
The large red circle has radius $R=30.0\arcsec$ (corresponding to $r\simeq 1.0$~kpc, at the distance of NGC~5474) and encloses the entirety of the optical bulge. The small cyan square marks the position of the kinematic center of the \HI~ disc, as determined by \citet[][R94, hereafter]{rownd} and confirmed by \citet[][K00, hereafter]{korn00}. It is worth noting that while the H$\alpha$ velocity field is compatible with having the same kinematic center as the \HI~, the amplitude of the H$\alpha$ rotation curve seems to be significantly larger 
\citep[$V_{max}{\rm sin(i})\simeq 25$~km/s vs. $V_{max}{\rm sin(i)}\simeq 8$~km/s;][]{epi,korn00}. Finally, the large blue circle is centred on the kinematic center of the disc and has a radius of $R=240.0\arcsec$, corresponding to $r\simeq 8.1$~kpc. \citet{korn00} trace the \HI~ disc out to $R\simeq 400\arcsec$, corresponding to $r\simeq 13.5$~kpc. It is apparent from the FUV and H$\alpha$ images of Fig.~\ref{foto2} that the youngest stars follow an irregular but clear and wide spiral pattern, whose center of symmetry roughly coincides with the kinematic center of the disc. However the strongest UV emission and some of the most remarkable \HII~regions are clustered in a bar-like structure that overlaps the bulge. Moreover the i-band image reveals the presence of a significant stellar over-density to the South-West of the bulge, with the shape of a fat crescent (SW over-density, see also Fig.~\ref{foto1}). 
The latter feature does not coincide with the spiral arms, but it seems to match the position of a local minimum in the \HI~distribution (see Fig.~9 by K00). The sub-structure is approximately comprised in the range of angular distances from the center of the bulge $50.0\arcsec \le R\le 110.0\arcsec$ (corresponding to the range 1.7~kpc to 3.7~kpc), with a surface density peak at $R\simeq 65\arcsec \simeq 2.2$~kpc (see also Fig.~\ref{hima}, below). 
It is interesting to note that the bulge does not seem strikingly off-set with respect to the overall spiral pattern. The strong displacement is with the kinematic center of the \HI ~distribution, and with the structure that, for historical reasons, we referred to as the ``bright disc'' (we will use again this name in the following, for brevity), that in fact is a nearly circular cloud of redder light (i.e., seen in i-band but not in FUV) that encloses/surrounds the SW over-density.

Both R94 and K00 find that the \HI ~disc displays a very regular velocity field in the central region, while
beyond $\simeq 180\arcsec$ from the kinematic center the typical features of a strongly warped disc become apparent, with the orientation of the kinematic major axis of the cold gas changing by more than $50\degr$.
The disc of the galaxy is nearly face-on with respect to the line of sight \citep[i$=21\degr$;][]{korn00}; this makes the derivation of the true amplitude of the rotation curve and, consequently, the estimates of the total dynamical mass, especially uncertain \citep[$M_{dyn}=2-6.5\times 10^9~M_{\sun}$;][]{rownd}.

%------------------------FIG -----------------------------------

%%%%%%%%%%%%%%%%%%%%%%%%%%%%%%%%%%%%%%%%%%%%%%%%%%%%%%%%%%%%%%%%%

The various dynamical disturbances observed in NGC~5474 are generally attributed to its interaction with M101, that also displays signs of interaction with the environment \citep[see, e.g.][]{mihos_OP} and is connected to NGC~5474 by a bridge of \HI~clouds and filaments \citep{HW79,mihos_HI}. However, as far as we know, a quantitative modeling of the interaction has never been attempted. 

Here we present the results of the analysis of the deep photometry of individual stars in NGC~5474 obtained and made publicly available by the LEGUS collaboration \citep{legus1,legus2}, from Hubble Space Telescope (HST) optical (F606W, F814W) images acquired with the Wide Field Channel (WFC) of the Advanced Camera for Surveys (ACS). In particular, for the first time, we reveal the stellar content of the various asymmetric components of the galaxy, providing new observational insight on their possible origin. This study is part of an ongoing  broader project aimed at searching for the observational signatures of the process of hierarchical merging in dwarf galaxies, the SSH survey \citep{ssh}.

Throughout the paper we adopt $(m-M)_0=29.22\pm 0.20$, corresponding to D=6.98~Mpc, from \citet{cflow2}, and foreground extinction $A_V=0.029$ from NED, in good agreement with \citet{galex}.
For the present-day oxygen abundance we assume  12+log(O/H)=$8.31 \pm 0.22$, estimated in \HII~regions by \citet{mousta} with the calibration by \citet{pili}, corresponding to $[M/H]\simeq -0.4$ and $Z\simeq 0.006$.

The outline of the paper is the following: in Sect.~\ref{sec_surf} we present newly derived light profiles of the bulge and of the nuclear star cluster; in Sect.~\ref{HST} we present the Color Magnitude Diagram and the spatial distribution of stars as a function of their age. Finally, in Sect.~\ref{discu} we discuss the main results of the analysis.

%\section{Data analysis}
%\label{sec_data}

%------------------------FIG -----------------------------------
   \begin{figure}
   \centering
   \includegraphics[width=\columnwidth]{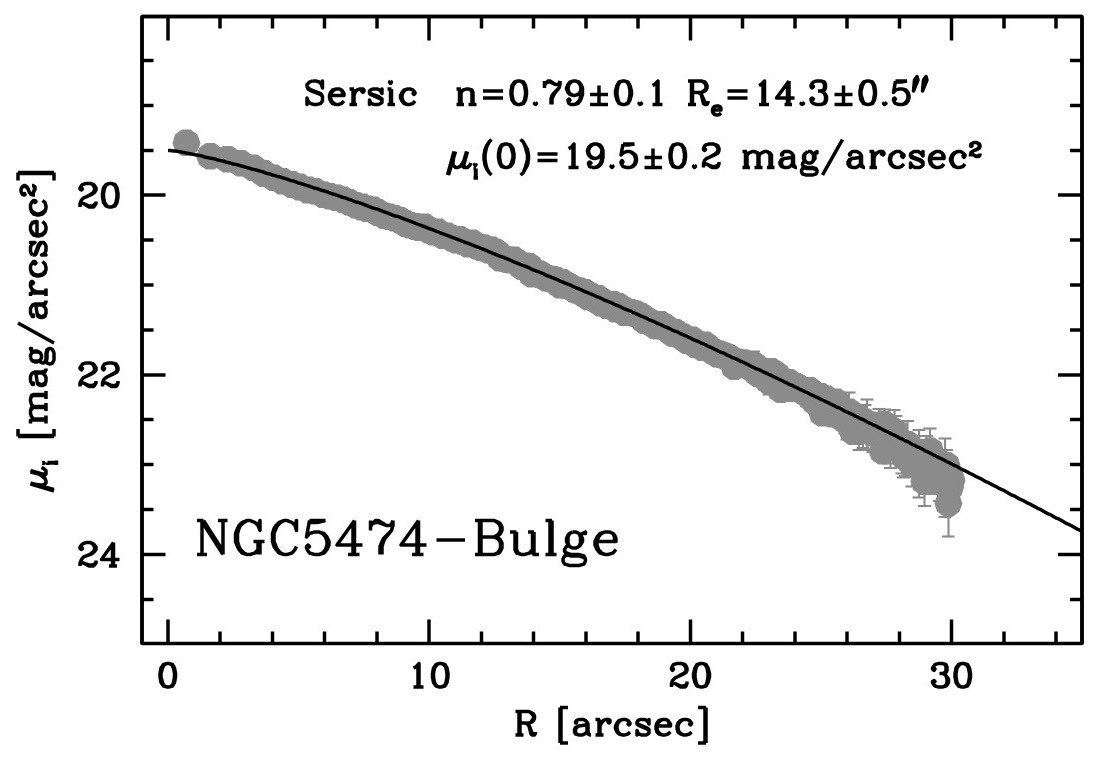}
     \caption{Surface Brightness profile of the bulge of NGC~5474 in the SDSS i band. 
     The continuous line is the S\'ersic model that best-fits the observed profile.
        \label{profi}}
    \end{figure}
%%%%%%%%%%%%%%%%%%%%%%%%%%%%%%%%%%%%%%%%%%%%%%%%%%%%%%%%%%%%%%%%%

%%%%%%%%%%%%%%%%%%%%%%%%%%%%%%%%%%%%%%%%%%%%%%%%%%%%%%%%%%%%%%%%%
\begin{table*}
  \begin{center}
  \caption{Properties of the bulge of NGC~5474}
  \label{bulge}
  \begin{tabular}{lccc}
\hline
name                        &  value   & units & note \\
\hline
%$(m-M)_0$                   &$29.22\pm 0.20$& mag  &from \citet{cflow2}    \\
%$A_V$                       & 0.029         & mag  &foreground extinction from NED     \\
%\hline
RA$_0$                      & 211.25587      & deg &position of the intensity peak \\
Dec$_0$                     &  53.66225      & deg &position of the intensity peak  \\
$r_{int}$                   &$12.78\pm0.1$& mag & from the best-fit S\'ersic profile in r-band$^a$     \\
$i_{int}$                   &$12.50\pm0.1$& mag  &from the best-fit S\'ersic profile in i-band$^a$     \\
$M_i$                       &$-16.7\pm 0.2$& mag &      \\
$M_V$                       &$-16.2\pm 0.3$& mag   & transformed from r and i magnitudes according to \citet{micche}     \\
$\langle\mu_i\rangle_e$     &$20.3\pm0.2$ & mag/arcsec$^2$ & from the best-fit S\'ersic profile in i-band$^a$    \\
$\langle\mu_V\rangle_e$     &$20.8\pm0.2$ & mag/arcsec$^2$& transformed from r and i magnitudes according to \citet{micche}     \\
$\mu_{V,0}$     &$20.1\pm0.2$ & mag/arcsec$^2$& transformed from r and i magnitudes according to \citet{micche}     \\
$R_e$                       & $14.3\pm 0.5$    &  arcsec&    \\
$r_e$                       & $484\pm20$    &   pc &   \\
$n$                         &$0.79\pm0.1$& &S\'ersic index     \\
$L_i$                       &$3.2\times10^8$& L$_{i,\sun}$$^b$ & \\
$M_{\star}$                 &$4.1\times10^8$& M$_{\sun}$$^c$ & \\
\hline
\multicolumn{4}{l}{$^a$ Using Eq.~9, Eq.~11 and Fig.~2 from \citet{GD05}, to pass from $\mu_e$ to 
$\langle\mu\rangle_e$ and to m$_{int}$.}\\
\multicolumn{4}{l}{$^b$ Adopting $M_{i,\sun}=4.57$ from { http://astroweb.case.edu/ssm/ASTR620/mags.html$\#$solarabsmag}}\\
\multicolumn{4}{l}{$^c$ Stellar mass obtained by adopting $M/L_{i,\sun}=1.27$. This is the mean of the values obtained for the $(r-i)_0$ color of the bulge from}\\ 
\multicolumn{4}{l}{the two relations provided by \citet{rc15}, fitted from two different sets of theoretical models.}\\
\end{tabular} 
\end{center}
\end{table*}
%%%%%%%%%%%%%%%%%%%%%%%%%%%%%%%%%%%%%%%%%%%%%%%%%%%%%%%%%%%%%%%%%

\section{Surface photometry}
\label{sec_surf}

To obtain structural parameters and estimate the stellar mass of the bulge we follow the same approach adopted by \citet{micche}, using r and i band images from SDSS DR12 \citep{dr12}. These sky-subtracted images are flux-calibrated in nMgy units\footnote{Nanomaggies, see\\ {\tiny\tt https://www.sdss.org/dr12/algorithms/magnitudes/}}, that can be easily converted into AB magnitudes with the relation mag=-2.5log(F[nMgy]+22.5. 

First we determined the center of the bulge as the position of the highest intensity peak on the image, with the aid of density contours. Then, the Aperture Photomety Tool software \citep[APT,][]{apt} was used to perform surface photometry. We derived the surface brightness profile over a series of concentric circular apertures covering the whole extension of the bulge, up to $33.5\arcsec$. 
To remove the contribution of the underlying disc, we estimated the background in a concentric annulus sampling that component and we subtracted it from the fluxes within the apertures. The background-subtracted surface brightness profile is shown in Fig.~\ref{profi}. 
All the parameters derived from the surface photometry are listed in Table~\ref{bulge}. We took the i band as a reference for the profile and for estimating the stellar mass because in this band the light contribution from young stars, presumably unrelated to the bulge, should be minimal.  
 
The parameters of the best-fit 
S\'ersic model \citep{sersic}\footnote{We adopt the formulation by \citet{cb99}, $\mu(R)=mu(0)+1.086b_n(R/R_e)^{\frac{1}{n}}$, where $R_e$ is the 2D half-light radius and $b_n$ is computed following \citet{ps97}.}, listed in Table~\ref{bulge}, are in good agreement with those obtained by \citet{fd10} from 3.6 $\mu$m band images from the Spitzer space mission, once the difference in the assumed distance are taken into account ($n=0.74\pm0.2$ and $r_e=439\pm 100$~pc, corrected to our distance). The structural parameters of the bulge of NGC~5474 are consistent with those of pseudo-bulges \citep{kk04,fd08}, and indeed it is classified as such by \citet{fd10}. It is interesting to note that these parameters are also fully compatible with the scaling laws of dwarf galaxies \citep[see, e.g.,][]{cote08,lange,micche,marchi}. In particular, the absolute integrated magnitude, the stellar mass, and the effective radius of the bulge are quite similar to those of NGC~205, a nucleated dwarf elliptical satellite of M~31 \citep[$M_V=-16.5\pm 0.1$, $M_{\star}=3.3\times10^8~M_{\sun}$, assuming $M/L_V=1.0$, and  $R_e=445\pm 25$~pc (circularised), from][]{mc12}.

%We estimated also the integrated r and i magnitudes of the ``bright disc'', including the SW over-density, by subtracting the flux of the bulge from the flux within an aperture of radius= $79.0\arcsec$, including the large majority of the light from this component. With the same assumptions as Tab.~\ref{bulge} we obtain $M_i=-17.7 \pm 0.3$, $(r-i)_0=-0.16\pm 0.05$, $M/L_{i,\sun}=0.08$ from \citet{rc15}, and, consequently, $M_{\star}=6.5\times10^7$~M$_{\sun}$. The stellar mass-to-light ratio and the stellar mass estimates are significantly more uncertain than those derived for the bulge, since magnitudes and colors of the disc are affected by a mix of populations that are widely different in age (see Sect.~\ref{HST}). 

\subsection{The stellar nucleus}
\label{sec_nucleus}

\begin{figure}
   \centering
   \includegraphics[width=\columnwidth]{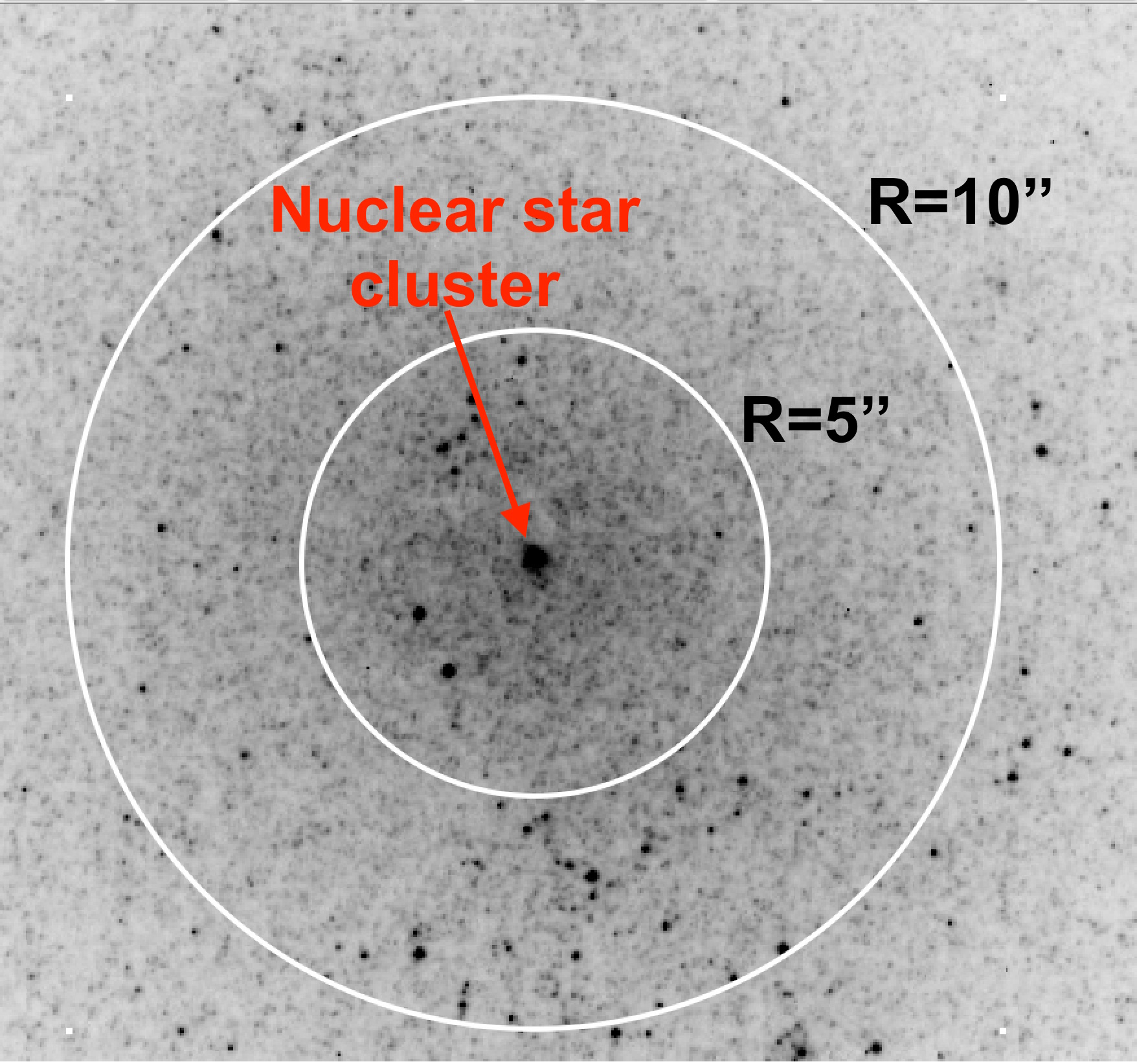}
   \includegraphics[width=\columnwidth]{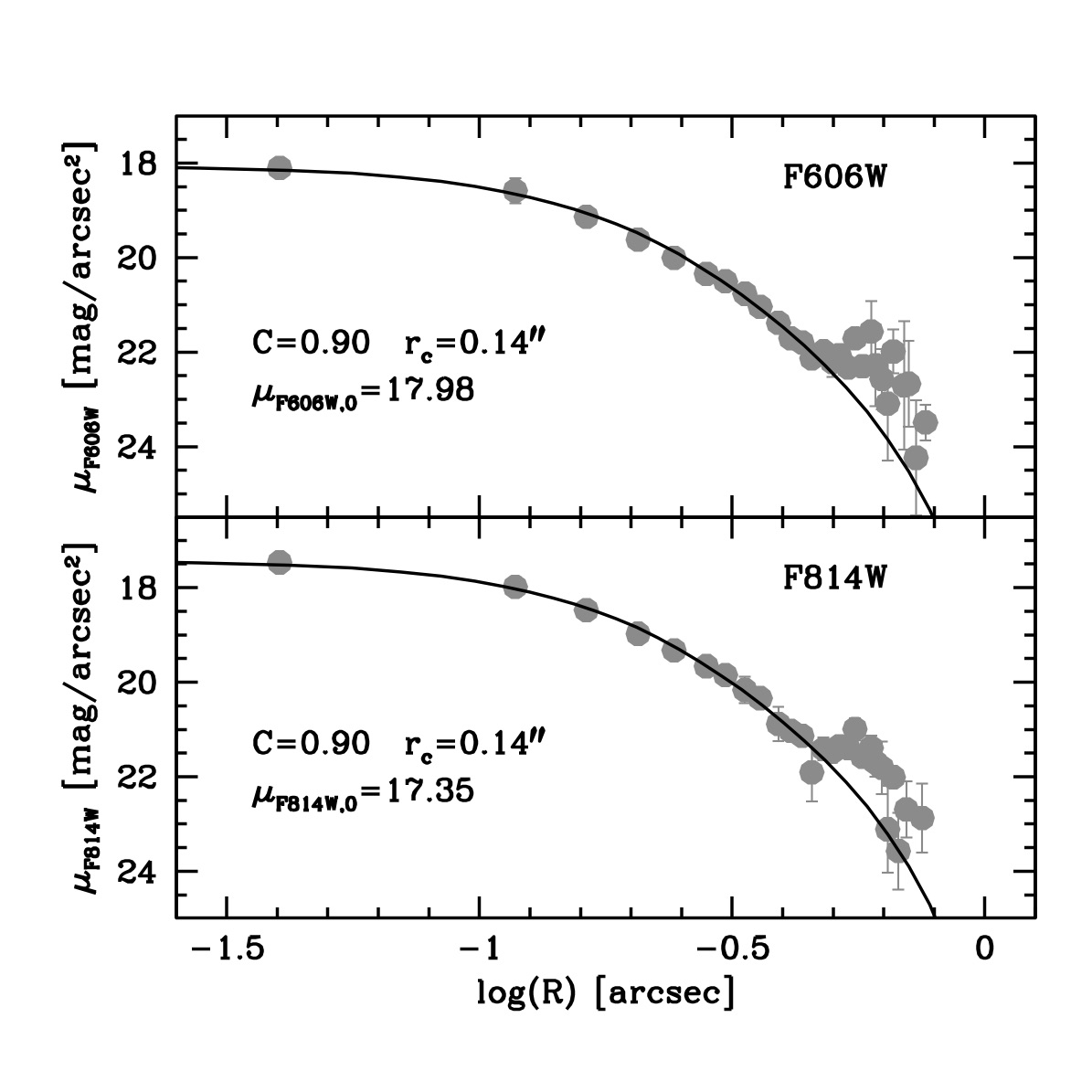}
     \caption{Upper panel: HST ACS F814W zoomed image of the central part of the bulge with intensity cuts chosen to make the stellar nucleus visible. The concentric circles provide the scale of the image and help the eye to appreciate that the nucleus reside at the center of the bulge.
     Lower panel: Surface Brightness profile of the stellar nucleus in both the ACS bands. 
     The continuous line displays the \citet{k66} model that fits the observed profiles. The parameters of the model are reported in the panels. The deviation from a smooth profile observed around log(R)=0.75 is due to a relatively bright source projected onto the outskirts of the nucleus.
        \label{nuc}}
    \end{figure}

The inspection of the ACS images revealed the presence of a star cluster residing at the center of the bulge (see the upper panel of Fig.~\ref{nuc}). The object is not resolved into stars but is clearly more extended than the stellar PSF. 

%------------------------FIG -----------------------------------
%%%%%%%%%%%%%%%%%%%%%%%%%%%%%%%%%%%%%%%%%%%%%%%%%%%%%%%%%%%%%%%%%

In the lower panels of Fig.~\ref{nuc} we show the surface brightness profile of the stellar nucleus obtained with APT aperture photometry on concentric circular annuli as above, from the  F606W and F814W ACS images. The profiles are nicely fitted by a \citet{k66} model with concentration $C=0.90$ and core radius $r_c=0.14\arcsec$. The theoretical profile has been convolved with a Gaussian distribution with FWHM=0.1$\arcsec$ to account for the smoothing due to the HST PSF. These and other measured properties of the system are listed in Table~\ref{nucleus}. The structural properties of the stellar nucleus derived here are in reasonable agreement with those provided by \citet{gb14}, who included it among those surveyed in their sample of 228 spiral galaxies.

%%%%%%%%%%%%%%%%%%%%%%%%%%%%%%%%%%%%%%%%%%%%%%%%%%%%%%%%%%%%%%%%%
\begin{table}
  \begin{center}
  \caption{Properties of the nuclear star cluster of NGC~5474}
  \label{nucleus}
  \begin{tabular}{lcc}
\hline
name                        &  value        & units  \\
\hline
RA$_0$                      & 211.2558      & deg\\
Dec$_0$                     &  53.6622      & deg\\
$F606W_{int}$               &$20.40\pm0.07$ & mag$^a$     \\
$F814W_{int}$               &$19.76\pm0.08$ & mag$^a$      \\
$M_V$                       &$-8.7\pm 0.2$  & mag$^b$   \\
$M_I$                       &$-9.5\pm 0.2$  & mag$^b$   \\
$(V-I)_0$                   &$0.75\pm 0.11$ & mag$^b$   \\
$\mu_{V,0}$                 &$18.1 \pm 0.1$ & mag \\
C                           &$0.90\pm 0.1$  &  \\
$r_c$                       &$0.14\pm 0.02$ & arcsec \\
$r_h$                       &$0.18\pm 0.02$ & arcsec$^c$ \\
$L_V$                       &$2.6\times10^5$& L$_{V,\sun}$  \\
$M_{\star}$                 &$2.6\times10^5$& M$_{\sun}$$^d$  \\
\hline
\multicolumn{3}{l}{$^a$ Aperture photometry.}\\
\multicolumn{3}{l}{$^b$ V and I magnitudes from \citet{har18} transformations.}\\
\multicolumn{3}{l}{$^c$ Projected half-light radius, from C and $r_c$, using Eq.~B4}\\ 
\multicolumn{3}{l}{of \citet{McL00}}\\ 
\multicolumn{3}{l}{$^d$ Adopting $M/L_V\simeq M/L_I= 1.0$, that is appropriate for the $(V-I)_0$}\\
\multicolumn{3}{l}{color of the nucleus according to, e.g., \citet{claudia} models.}\\
\end{tabular} 
\end{center}
\end{table}
%%%%%%%%%%%%%%%%%%%%%%%%%%%%%%%%%%%%%%%%%%%%%%%%%%%%%%%%%%%%%%%%%

The nuclear cluster has  a stellar mass typical of globular clusters (GCs) and a $(V-I)_0$ color compatible with the most metal-poor Galactic GCs \citep{ss07}. It fits well the relation between $r_h$ and $C$ of Galactic globular clusters \citep[GGC,][]{djo03}, but its concentration parameter is lower than any GGC with similar $M_V$ \citep{dm94}. The ratio of its stellar mass to that of the bulge as a whole obeys the relation that is known to exist between the stellar masses of galaxies and of their nuclei \citep[see, e.g.,][and references therein]{sanchez}. The presence of a nuclear star cluster is a further element of similarity with NGC~205 \citep[see, e.g.,][and references therein]{heat}; the color of the nuclei are also similar \citep{ngu}.

\section{Photometry of resolved stars}
\label{HST}

\begin{figure}
   \centering
   \includegraphics[width=\columnwidth]{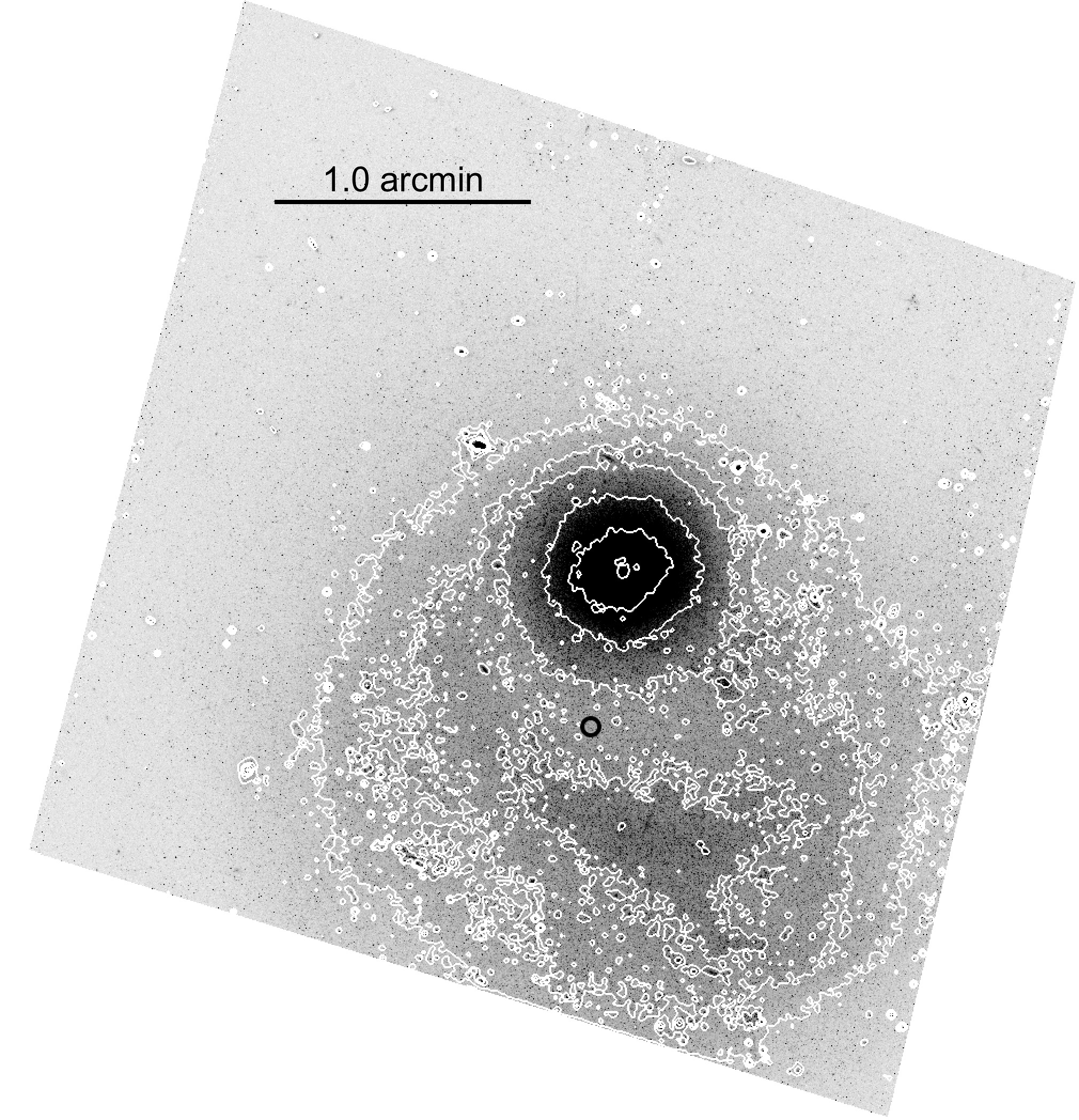}
     \caption{HST ACS drizzled image of NGC~5474 in the F814W filter. The intensity contours are at 0.04, 0.08, 0.12, 0.25, 0.5, 1.0, 2.0 $e^{-}/s$. The black circle marks the position of the kinematic center of the \HI~ disc.
        \label{hima}}
    \end{figure}

We retrieved\footnote{From the LEGUS web site {\tt https://legus.stsci.edu}} the F606W and F814W photometric catalogs produced by LEGUS \citep{legus1} from ACS images nearly centred on the bulge of NGC~5474. The F814W drizzled image at the origin of this dataset is shown if Fig.~\ref{hima}. In this figure the intensity contours highlight very clearly the bulge and the SW over-density. 
All the details on the data reduction and on the parameters included in the catalogs are reported in \citet{legus2}.  

First we selected from each catalog only the sources classified as bona fide stars, then we applied further selections based on various quality parameters. In particular we required 
that stars retained in the final catalogs satisfy the following conditions: {\tt ROUND}$< 3.0$, {\tt CHISQ}$<2.0$, {\tt |SHARP|}$<0.2$, and {\tt CROWD}$<0.2$, for both F606W and F814W \citep[see][for the meaning of the parameters]{legus2}. We matched the two catalogues using CataXcorr\footnote{\tt \tiny http://davide2.bo.astro.it/\~ ~paolo/Main/CataPack.html} with a first degree transformation, ending up with a merged catalog with 137174 stars. Finally, we added to our catalog V,I magnitudes in the Johnson-Cousins system, obtained from F606W and F814W magnitudes with the transformations by \citet{har18}. 

%------------------------FIG -----------------------------------

%%%%%%%%%%%%%%%%%%%%%%%%%%%%%%%%%%%%%%%%%%%%%%%%%%%%%%%%%%%%%%%%%

We transferred to the catalog the astrometry embedded in the drizzled images provided by LEGUS. This was done by reducing the F606W image with Sextractor \citep{sex}, with a $10\sigma$ detection threshold, to obtain an output catalog with positions in RA and Dec. Then we used CataXcorr to cross-correlate this catalog with the catalog we got from LEGUS, and we transformed pixel coordinates into equatorial coordinates with a first degree polynomial in x,y fitted on 32571 stars in common, well distributed over the whole field. The residual of the transformation are $<0.2\arcsec$ in both coordinates. The astrometric solution embedded in drizzled images may have a small zero-point residual. We checked with external catalogues and we concluded that the residual should be $\la 1.0\arcsec$, that is negligible for the purposes of the following analysis.
 
%------------------------FIG -----------------------------------
   \begin{figure}
   \centering
   \includegraphics[width=\columnwidth]{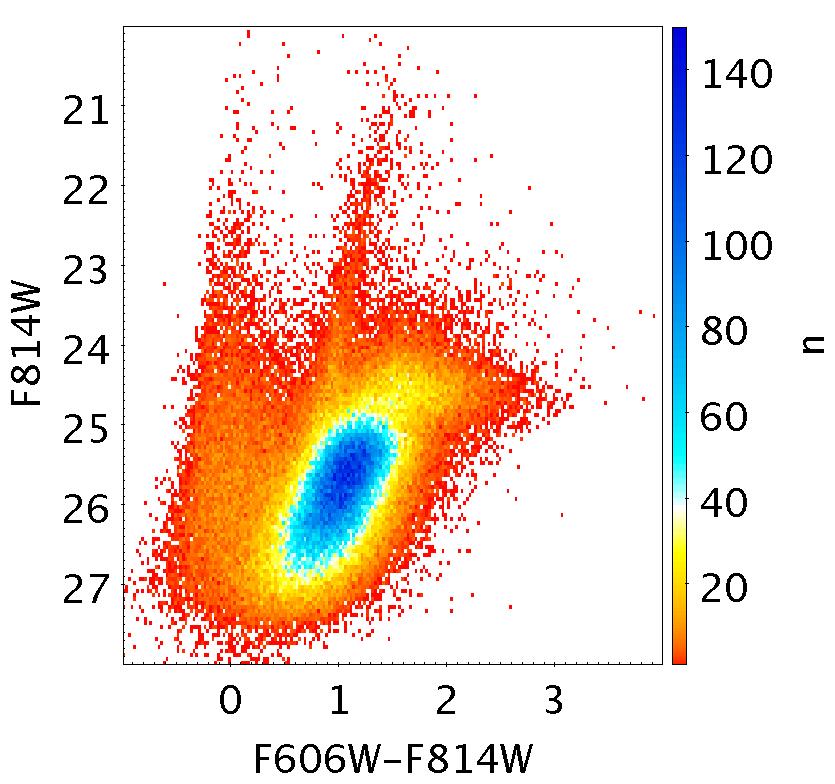}
   \includegraphics[width=\columnwidth]{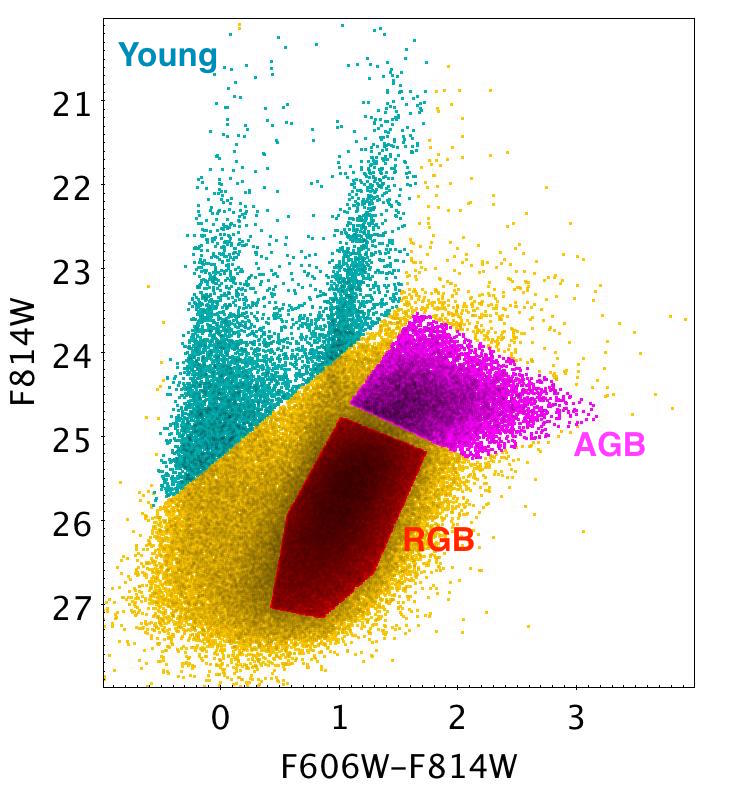}
     \caption{Upper panel: CMD from the HST catalog. 
     %The color is proportional to the square root of the local density. 
     Lower panel: CMD illustrating the boxes adopted to select the various populations.
        \label{cmd}}
    \end{figure}

%%%%%%%%%%%%%%%%%%%%%%%%%%%%%%%%%%%%%%%%%%%%%%%%%%%%%%%%%%%%%%%%%

\subsection{The Color Magnitude Diagram}
\label{cmdsec}

The upper panel of Fig.~\ref{cmd} shows the Color Magnitude Diagram (CMD) of the entire sample of selected stars in the ACS field, color coded according to the square root of the local density. Most of the stars are tightly clustered into a well defined Red Giant Branch (RGB), running from F606W-F814W$\simeq 0.5$ at F814W$\simeq 27.0$ to a clear tip located at F814W$\simeq 25.1$ and $\langle {\rm F606W-F814W}\rangle\simeq 1.1$. Stars in this evolutionary phase are older than 1-2~Gyr, but possibly as old as $> 10$~Gyr. Above the RGB tip a broad sequence of bright stars bends to the red and culminates at F814W$\sim 24.3$. These are Asymptotic Giant Branch (AGB) stars, tracing intermediate to old age populations (age$\ga 100$~Myr). The blue vertical plume at F606W-F814W$\la 0.3$ hosts both Main Sequence stars and stars at the blue edge of the core Helium burning phase. The thin diagonal sequence emerging 
at the RGB tip level, to the blue of the RGB and AGB sequences, and running up to F814W$\simeq 20.0$ at  F606W-F814W$\simeq 1.3$ (red plume) collects stars located at the red edge of the core Helium burning phase. All the stars lying in the blue and red plumes and between them are younger than a few $\simeq 100$~Myr. 

In the lower panel of Fig.~\ref{cmd} we illustrate the selection boxes we adopt to trace stellar populations in different age ranges: RGB, AGB and Young stars from both the blue and red plumes. A comparison with theoretical isochrones from the PARSEC set \citep{parsec} allows a better age bracketing of the stars enclosed in our selection boxes, taking into account also the effect of metallicity. In particular, using isochrones with metallicity corresponding to the oxygen abundance measured in the disc of NGC~5474 (Z=0.006) we find that our Young selection box is populated only by stars younger than 100~Myr. The sharp break in the distribution of AGB stars at F814W$\sim 24.3$ is matched by isochrones of age $\ga 630$~Myr, for Z=0.006, and of age $\ga 1.0$~Gyr for Z=0.0006 (stars significantly more metal-poor than the present day metallicity are likely part of the stellar mix, when old-intermediate age populations are considered). We conclude that our AGB box includes stars in the age range from $\simeq 0.5$~Gyr to $4-5$~Gyr, since, according to the adopted stellar evolution models, older age AGB stars in this range of metallicity do not become brighter than the RGB tip. Finally, under the same assumptions, selected RGB stars  should have ages in the range $\simeq 2.5$~Gyr to $\simeq 12.5$~Gyr. In the following section we will show and discuss the radial distribution of the various tracers.

%------------------------FIG -----------------------------------
%   \begin{figure}
%   \centering
%   \includegraphics[width=\columnwidth]{comple.jpg}
%     \caption{F814W faint magnitude limit that includes 50$\%$ (f=0.50, empty triangles) 
%     and 90$\%$ (f=0.90, empty squares) of the stars having $0.4<F606W-F814W<3.5$ and $F814W>23.0$, as a function of the angular distance from the center of the bulge. These parameters traces the radial variation of the completeness in the sample. The variation is very strong in the region of the bulge while it is mild for $R> 30.0\arcsec$.
%        \label{comple}}
%    \end{figure}

%%%%%%%%%%%%%%%%%%%%%%%%%%%%%%%%%%%%%%%%%%%%%%%%%%%%%%%%%%%%%%%%%
%------------------------FIG -----------------------------------
   \begin{figure*}
   \centering
   \includegraphics[width=\textwidth]{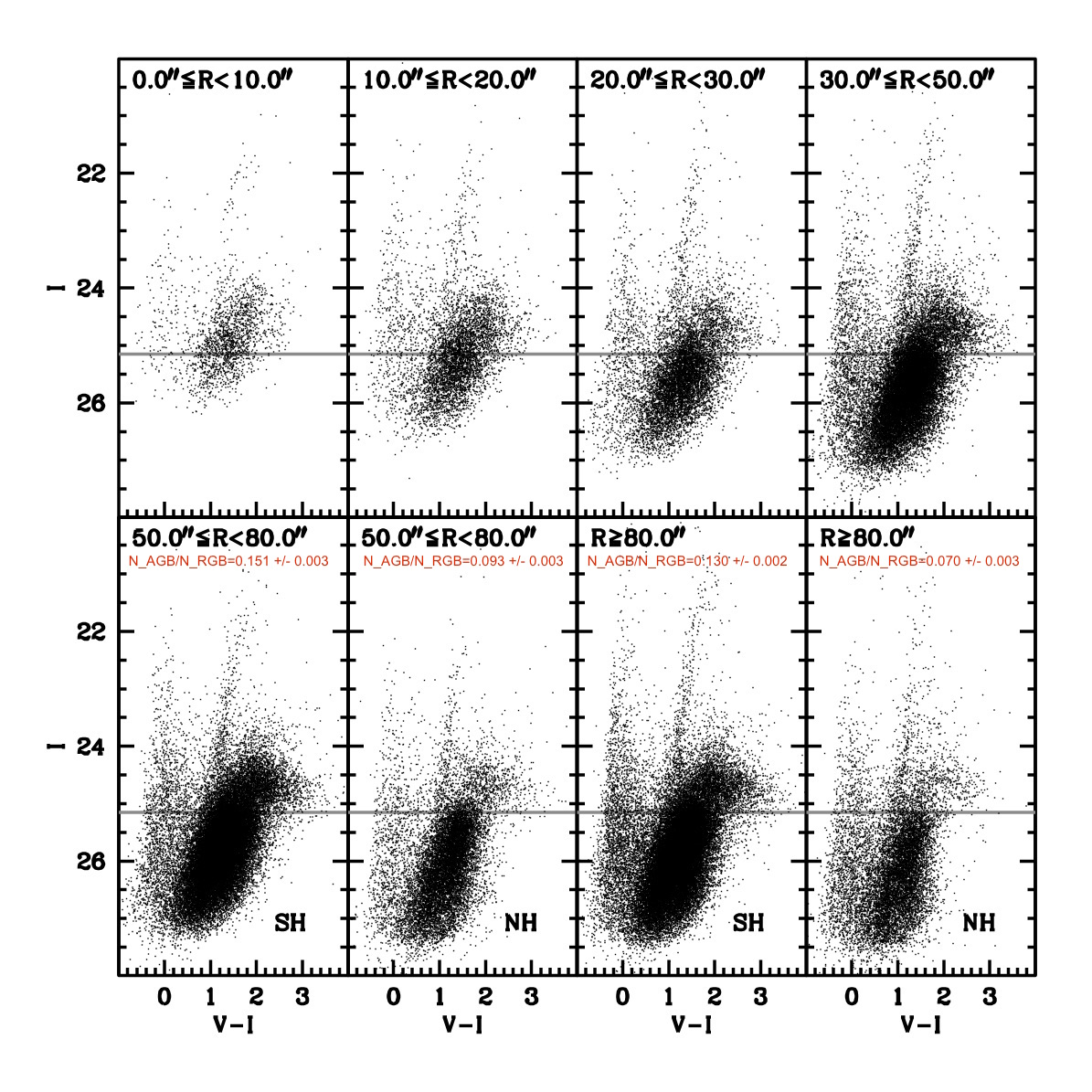}
     \caption{CMD in different radial annuli, from the center of the bulge to the edge of the ACS image. For $R\ge 50.0\arcsec$ we split the annuli in their Northern and Southern halves (NH and SH, respectively), to highlight the asymmetry in the surface density related to the SW over-density. In the corresponding panels we report also the ratio between AGB and RGB stars. The horizontal grey line marks the position of the RGB tip as determined from the whole sample. This series of CMDs also clearly illustrate the severe effect of the increasing incompleteness toward the center of the most crowded substructure, i.e. the bulge, with the limiting magnitude dropping by nearly 2~mag from $R>30.0\arcsec$ to $R<10.0\arcsec$.
        \label{radial}}
    \end{figure*}

%%%%%%%%%%%%%%%%%%%%%%%%%%%%%%%%%%%%%%%%%%%%%%%%%%%%%%%%%%%%%%%%%

In Figure~\ref{radial}, we show the CMD for different radial annuli centred on the photometric center of the bulge. This figure illustrates the strong impact of radially varying incompleteness in the region enclosing the bulge ($R<30.0\arcsec$), with the limiting magnitude dropping from F814W$\simeq 27.0$ for $R\ge 30.0\arcsec$ to
F814W$\simeq 25.5$ for $R< 10.0\arcsec$. In the innermost annuli we notice the smearing of the main CMD features due to the effects of stellar blending and to the increased photometric uncertainty related to the extreme degree of crowding in the bulge region. Outside $R=30.0\arcsec$ the quality and completeness of the photometry is nearly constant. Fig.~\ref{radial} also shows that the same evolutionary sequences are present everywhere in the considered field, albeit not necessarily with the same relative abundance. The two outermost annuli have been split into their Southern and Nothern halves (SH and NH, respectively). In these radial ranges ($R\ge 50.0\arcsec$) the ``bright disc'', and, in particular, the SW over-density, are completely included in the SH, while the NH lack any obvious structure (see Fig.~\ref{hima}; in particular the densest part of SW is included in the SH annulus 
$50\arcsec \le R\le 80.0\arcsec$). This difference is clearly reflected in the corresponding CMDs. However, in spite of the strong asymmetry in favour of the SH annuli, a significant number of stars associated to NGC~5474 is present in the NH also in the outermost regions, for $R\ge 80.0\arcsec$. Hence, an extended distribution of intermediate age and old stars (a halo component?) is present everywhere in the sampled field.

It is interesting to note that the ratio of the number of AGB and RGB stars is significantly larger in SH than in NH, suggesting that intermediate age stars represent a larger fraction of the old+intermediate age population in the region to the South of the bulge, dominated by the SW over-density, than in the ``empty'' region to the North of it. While the actual values of this ratio are affected by the incompleteness of the RGB sample (that reaches the limiting magnitude of the photometry) this should affect equally the northern and southern halves of the same radial annulus, thus making the comparison between the ratios in SH and NH meaningful. We have verified that the reported differences are unchanged if only RGB stars brighter than $I=26.0$ are used, a subset much less affected by incompleteness.

\subsection{The spatial distribution of stars}
\label{distri}

\begin{figure}
   \centering
   \includegraphics[width=\columnwidth]{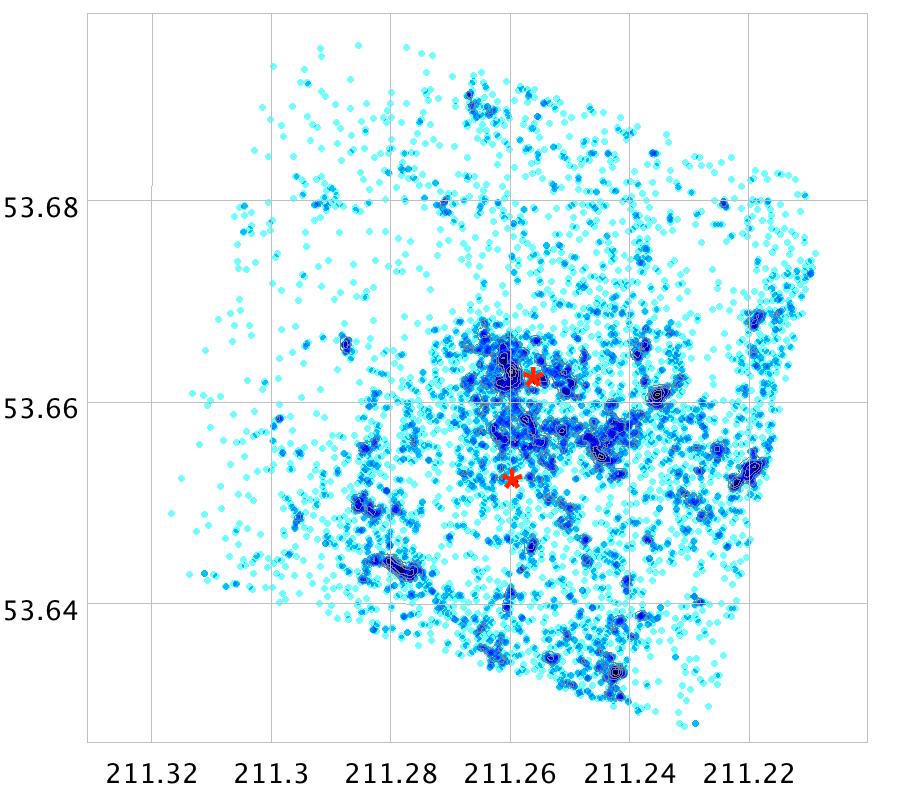}
     \caption{Density map of the YOUNG sample obtained with TOPCAT. The axis are RA and Dec (in degrees) in sin projection. North is up and East is to the left. The color is proportional to the square root of the local density, darker tones of blue traces higher density levels. The two asterisk symbols superimposed to the map mark the position of the center of the bulge and of the dynamical center of the \HI ~disc \citep[from][]{rownd}, from North to South, respectively.
        \label{young}}
    \end{figure}

\begin{figure}
   \centering
   \includegraphics[width=\columnwidth]{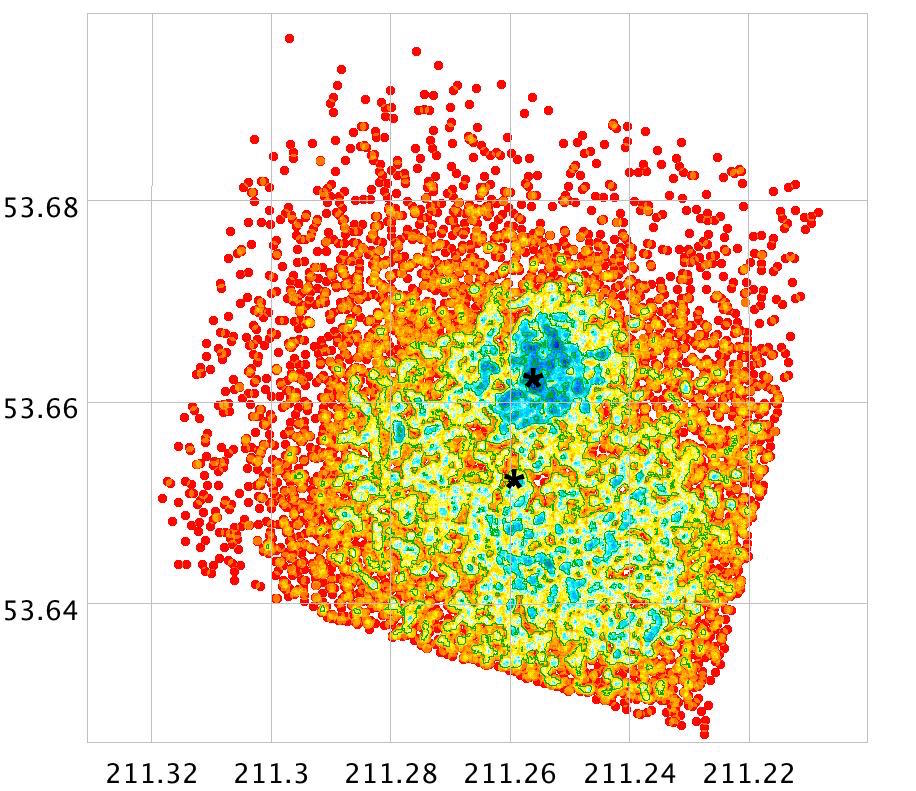}
     \caption{Same as Fig.~\ref{young} for the AGB sample. Here the color code goes from warm colors (low density) to cold colors (high density).
        \label{agb}}
    \end{figure}

\begin{figure}
   \centering
   \includegraphics[width=\columnwidth]{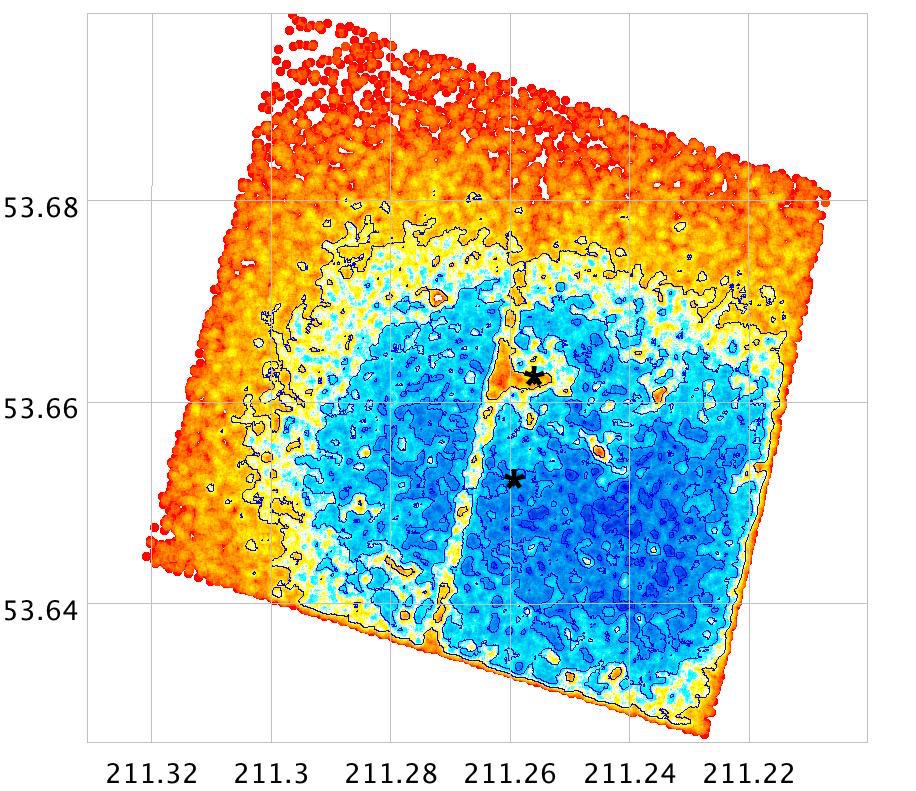}
     \caption{Same as Fig.~\ref{agb} for the RGB sample. 
        \label{rgb}}
    \end{figure}

The density maps of the stars belonging to the CMD selection boxes described above are shown in Figs.~\ref{young}, \ref{agb}, and \ref{rgb}, for the YOUNG, the AGB, and the RGB samples, respectively. The positions of the photometric center of the Bulge and of the kinematic center of the \HI ~disc are reported in all the maps
(upper and lower asterisk, respectively).

%------------------------FIG -----------------------------------

%%%%%%%%%%%%%%%%%%%%%%%%%%%%%%%%%%%%%%%%%%%%%%%%%%%%%%%%%%%%%%%%%

%------------------------FIG -----------------------------------

%%%%%%%%%%%%%%%%%%%%%%%%%%%%%%%%%%%%%%%%%%%%%%%%%%%%%%%%%%%%%%%%%

%------------------------FIG -----------------------------------

%%%%%%%%%%%%%%%%%%%%%%%%%%%%%%%%%%%%%%%%%%%%%%%%%%%%%%%%%%%%%%%%%

The stars younger than $\simeq 100$~Myr (YOUNG) clearly display the same spiral pattern shown in the left panel of Fig.~\ref{foto2}, including the central bar-like structure \citep[see also][for an analysis of the youngest stars]{cignoni}. While the pattern is visible over the whole field, the densest clusters of young stars lie in the South-West half, showing that also the recent star formation has been asymmetric on large scales (a few kpc). 

The intermediate-age AGB stars have a very different distribution: the bulge and the SW over-density stand out as the most remarkable structures, possibly connected by a fainter bridge of stars protruding from the eastern edge of the over-density approximately toward the East and reaching the eastern edge of the bulge, after a strong bending toward North-West.

\begin{figure}
   \centering
   \includegraphics[width=\columnwidth]{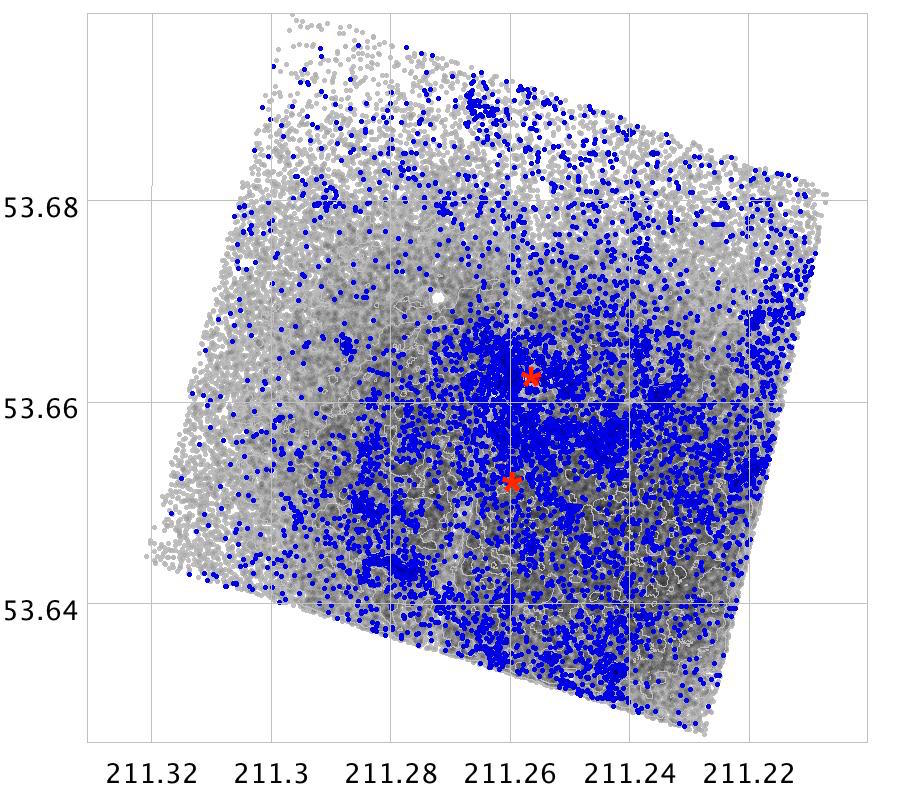}
     \caption{Stars of the YOUNG sample (in blue) are superposed to stars of the RGB sample (in grey) in a map with the same arrangement and symbols as in Fig.~\ref{rgb}. 
        \label{young_rgb}}
    \end{figure}

The map of old RGB stars (Fig.~\ref{rgb}) displays a density minimum at the position of the bulge. We should emphasise, however, that the lack of RGB stars in the central bulge is entirely due to the strong incompleteness affecting this region (see Sect.~\ref{HST}) and thus preventing the detection of stars fainter than the AGB. However, the CMD of the outer corona of the bulge (Fig.~\ref{radial}) strongly suggest that RGB stars are an important component of the bulge, albeit unresolved near the center.
On the other hand, the SW over-density and the stellar bridge identified on the AGB map are quite evident also in this map.  Hence, we conclude that old and intermediate-age stars traces the SW over-density, dominating its stellar content.  On the other hand, Fig.~\ref{young_rgb}, where the density map of the YOUNG stars is superimposed to that of RGB stars shows that the main spiral arms and the SW over-density are not spatially correlated. 

 %------------------------FIG -----------------------------------

%%%%%%%%%%%%%%%%%%%%%%%%%%%%%%%%%%%%%%%%%%%%%%%%%%%%%%%%%%%%%%%%%

\section{Discussion and Conclusions}
\label{discu}

Large asymmetries in the distribution of stars in a galaxy like NGC~5474 can be produced only by (a) inherently asymmetric star formation episodes, that are quite frequent in star forming galaxies, or by (b) a strong dynamical disturbance, due to an interaction with another galaxy. 

The old age ($> 2$~Gyr) of the stars populating the SW over-density militates against the first hypothesis, as orbital mixing should wipe out an over-density originated by an asymmetric episode of star formation within the disc of NGC~5474 in a few orbital times. However this possibility cannot be completely dismissed, since the SW over-density lies in the rising branch of the rotation curve, the rise is approximately solid body \citep{epi} and the orbital periods at the relevant distances from the \HI ~kinematic center are relatively long (in the range 200-400~Myr assuming the \citealt{epi} velocity curve, and $>500$~Myr adopting the curve by \citealt{korn00}).

A major episode of star formation producing such a conspicuous and long-lasting substructure may have also been triggered by dynamical interaction, the generally hypothesised past encounter with M~101 being the most obvious candidate. However, assuming a difference in transverse velocity between M~101 and NGC~5474 as large as 300~km/s (about six times the difference in radial velocity) and that the projected separation (90 kpc) corresponds to the 3D mutual distance, the close encounter with M101 should have occurred  $\sim 300$~Myr ago, 
while the hypothetic burst started more than 2~Gyr ago. 

In conclusion the variety of anomalies observed in NGC~5474 suggests that it is advisable to consider also the possibility that, in addition to the interaction with M~101, the galaxy bears also the signs of a recent interaction with a dwarf companion. For instance, the SW over-density could be the remnant of a recently disrupted satellite, whose merging triggered the observed enhancement in the recent star formation in the Southern half of the gaseous disc. Alternatively, it can be the density wake produced by the off-set of the bulge within the disc or by the passage of a companion. Long-lasting (up to $\simeq 1$~Gyr) spiral patterns induced by impulsive interactions have been observed in N-body simulation of galaxy encounters \citep[see, e.g.,][and  references therein]{struck}. Still, at least in the cases explored by \citet{struck}, they show up as two-arms spiral patterns in stars and gas, that do not seem to match the distribution of old and intermediate-age stars shown in Fig.~\ref{agb} and Fig.~\ref{rgb}.

A natural candidate as a still-alive companion is the putative bulge, an hypothesis already mentioned by other authors \citep[e.g.,][]{mihos_OP}. Indeed, \citet{mihos_OP} concluded that the relatively regular outer isophotes of the galaxy are not easy to reconcile with a strong tidal interaction, e.g., with a massive galaxy like M~101, and that the off-centered position of the bulge is suggestive of a recent or on-going interaction with a less conspicuous body. 

Some broad consistency check of this hypothesis can be attained, based on the structural similarity of the bulge with local dwarfs (Sect.~\ref{sec_surf}). 
The latter obey to a tight relation between $M_V$ and dynamical mass $M_{dyn}$, or, equivalently, the dynamical mass to light ratio $M_{dyn}/L$, with $M_{dyn}/L$ in solar units ranging from $\sim 1000$ for the faintest dwarfs to
$\sim 1-3$ for the brightest ones, with luminosities similar to the bulge of NGC~5474 \citep[where all the masses and mass to light ratios are computed within the half light radius,][]{mc12}. According to this author, the local galaxy that is more similar to it, NGC~205 (see Sect.~\ref{sec_surf}), has $M_{dyn}/L\simeq M_{dyn}/M_{star} \simeq 2$. If we adopt the same for the bulge as a whole we obtain $M_{dyn}\sim 8\times 10^8$~M$_{\sun}$, a factor of $\sim 2.5-8$ lower than the dynamical mass of NGC~5474, as estimated from the rotation curve of its gaseous disk \citep[$2.0-6.5\times 10^9$~M$_{\sun}$, within 8.8~kpc from the center;][]{rownd}. A mass ratio $\sim 0.1-0.2$ would probably be broadly consistent with visible but not destructive effects of the encounter. In this case NGC~5474 would be a bulge-less disc galaxy with a minor stellar halo component and a dynamical mass-to-light ratio of $(M/L_i)_{dyn}\sim 20-60$. At a first glance, and given the available data, this scenario looks plausible, or, at least, worth of further analysis. 

If the off-set bulge is not actually a bulge but, instead, an early-type satellite orbiting around the disc of NGC~5474, then it should have a systemic line of sight velocity slightly different from that of the disc, at the same location. If the hypothetic pair is bound, the difference in the systemic velocity along the line of sight should amount to a few tens of km/s, at most. Unfortunately the available stellar velocity fields (or individual long slit measures) are all based on emission lines, tracing the kinematics of the star-forming disc, also in the central regions \citep{ho95,mousta,epi}. On the other hand, the bulge is dominated by old and intermediate-age stars, hence its kinematics can be traced only with absorption line spectra, that are missing\footnote{Stellar velocity fields from both emission and absorption lines may be possibly obtained from the data described in \citep[][PINGS project]{rosa10}. However the results for NGC~5474 have been not presented yet.}.
In principle one can obtain a spectrum of the bulge region including both emission lines (from the disc) and absorption lines (from the bulge) and check if the radial velocity of the two components is the same or not.

We did try to perform this test thanks to one night of Director Discretionary Time kindly made available at the 3.5m Telescopio  Nazionale Galileo (TNG) in La Palma. Unfortunately the quality of the data and the reliability of their wavelength calibration were not sufficient to reach our scientific goal (see Appendix~\ref{appA} for a description of the encountered problems). The only conclusion that we can draw from the analysis of the acquired spectra is that the radial velocity difference between emission and absorption lines within the bulge is $\le 50$~km/s, thus leaving the key question on the nature of the bulge unanswered. However both the limit in the velocity difference between the disc and the bulge and the similarity of the CMD of the two components, indicating similar distances, rules out the hypothesis of the chance superposition of unrelated systems, suggested, e.g., by \citet{rownd}.

In summary, the present analysis provides an important piece of information that was lacking from the overall picture of this highly disturbed galaxy, that is the actual stellar contentASTROPH and age distribution of the various and structures and substructures. This is an important element to
set the scene for a serious attempt to reproduce the main properties of NGC~5474 with dynamical (and/or hydrodynamical) simulations, a task that we are planning for the near future.

%%%%%%%%%%%%%%%%%%%%%%%%%%%%%%%%%%%%%%%%%%%%%%%%%%%%%%%%%%%%%%%%%
\begin{acknowledgements}

We are grateful to the Referee, Curtis Struck, for a very careful reading of the manuscript and for the useful suggestions for the next steps of the analysis of the NGC~5474 + M~101 system.

This research is partially funded through the INAF Main Stream program ``SSH'' 1.05.01.86.28. F. A., M.C. and M.T. acknowledge funding from INAF PRIN-SKA-2017 program 1.05.01.88.04.

Based on observations made with the NASA/ESA Hubble Space Telescope, obtained at the Space Telescope Science Institute, which is operated by the Association of Universities for Research in Astronomy, under NASA Contract NAS 5-26555. These observations are associated with Program 13364.

Based on observations made with the Italian Telescopio Nazionale Galileo (TNG) operated on the island of La Palma by the Fundación Galileo Galilei of the INAF (Istituto Nazionale di Astrofisica) at the Spanish Observatorio del Roque de los Muchachos of the Instituto de Astrofisica de Canarias". We are grateful to the TNG staff for their support in preparing the observations. Special thanks are due to the TNG Director Ennio Poretti, to Walter Boschin,  Antonio Magazz\`u and Luca Di Fabrizio.

This research made use of SDSS data. Funding for the SDSS and SDSS- II has been provided by the Alfred P. Sloan Foundation, the Participating Institutions, the National Science Foundation, the US Department of Energy, the National Aeronautics and Space Administration, the Japanese Monbukagakusho, the Max Planck Society, and the Higher Education Funding Council for England. The SDSS Web Site is {\tt http://www.sdss.org}. The SDSS is managed by the Astrophysical Research Consortium for the Participating Institutions. The Participating Institutions are the American Museum of Natural History, Astrophysical Institute Potsdam, University of Basel, University of Cambridge, Case Western Reserve University, University of Chicago, Drexel University, Fermilab, the Institute for Advanced Study, the Japan Participation Group, Johns Hopkins University, the Joint Institute for Nuclear Astrophysics, the Kavli Institute for Particle Astrophysics and Cosmology, the Korean Scientist Group, the Chinese Academy of Sciences (LAMOST), Los Alamos National Laboratory, the Max-Planck-Institute for Astronomy (MPIA), the Max-Planck- Institute for Astrophysics (MPA), New Mexico State University, Ohio State University, University of Pittsburgh, University of Portsmouth, Princeton University, the United States Naval Observatory, and the University of Washington.

Most of the analysis presented in this paper has been performed with TOPCAT \citep{topcat}.
This research has made use of the SIMBAD database, operated at CDS, Strasbourg, France.
This research has made use of the NASA/IPAC Extragalactic Database (NED) which is operated by the Jet Propulsion Laboratory, California Institute of Technology, under contract with the National Aeronautics and Space Administration. 
This research has made use of NASA's Astrophysics Data System.

\end{acknowledgements}

\bibliographystyle{apj}

\appendix

\section{Spectroscopy}
\label{appA}

In an attempt to perform the test described in Sect.~\ref{discu} we obtained one night of Director Discretionary Time at the 3.5~m Telescopio Nazionale Galileo (TNG), located at the International Observatory Roque de los Muchachos (La Palma, Spain).
We used the Dolores spectrograph in long-slit mode, with the V510 grism that covers the wavelength range
4875\AA - 5325\AA. To improve the light collection we adopted a slit width of $2.0\arcsec$, implying a spectral resolution of $R_{\lambda}\sim 3000$, that, in principle, should lead to uncertainties $\la 10.0$~km/s in the radial velocity (RV) estimates. If the ``bulge'' is a satellite of NGC~5474, the velocity difference between the two should be of the order of the amplitude of the rotation curve, i.e. $\la 60$~km/s \citep[][corrected for the inclination of the disc]{epi}. 

%------------------------FIG -----------------------------------
%   \begin{figure*}
   \begin{figure}
   \centering
  \includegraphics[width=\columnwidth]{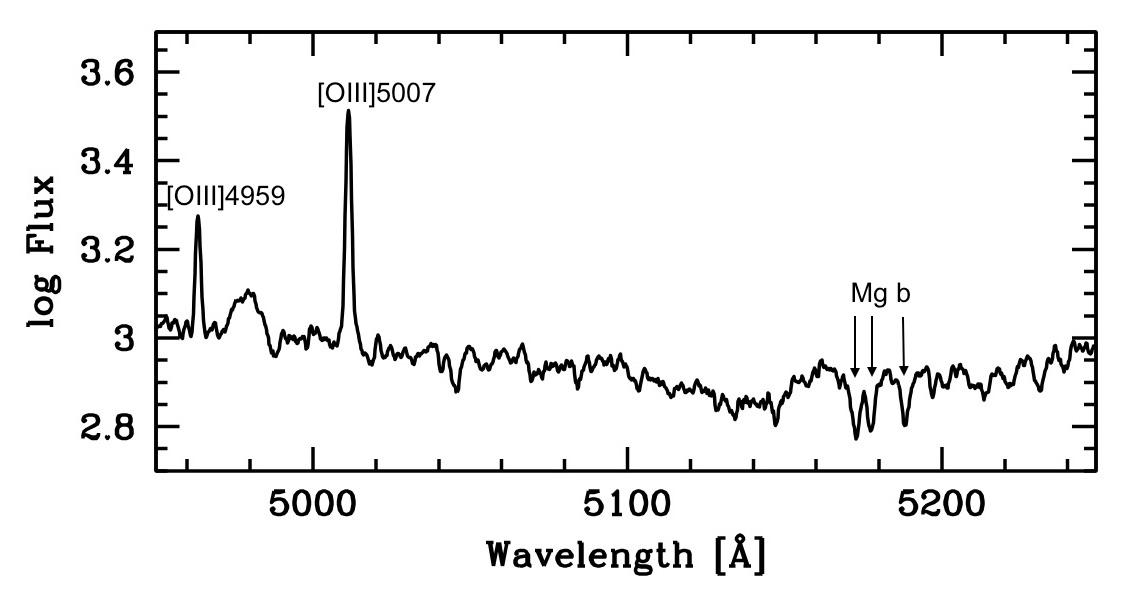}
     \caption{Stacked spectrum of the center of the NGC~5474 bulge, smoothed with a 5~px boxcar filter.  The most prominent emission and absorption lines are labeled. The flux scale is in arbitrary units.
     %The wavelength scale factor is $\simeq 0.22$~\AA/px, the pixel angular size on sky is $0.252\arcsec$.
        \label{vedis}}
    \end{figure}
%    \end{figure*}

%%%%%%%%%%%%%%%%%%%%%%%%%%%%%%%%%%%%%%%%%%%%%%%%%%%%%%%%%%%%%%%%%

Observations were performed in service mode during the night of April 8, 2019. Six $t_{exp}= 3150$~s on-target spectra were acquired. To minimise the systematic effects due to instrument flexures, a wavelength calibration lamp should have been acquired before and after each science exposure. Unfortunately it was impossible to follow this optimal practice because: (a) the overall procedure to acquire the spectrum of a Thorium-Argon calibration lamp is very expensive in term of observing time ($\simeq 1800$~s), and, (b) the introduction of the lamp in the light path implies the interruption of the telescope guiding. Therefore, after each lamp spectrum the target should be newly acquired and the previous position of the slit cannot be exactly reproduced. For this reason, as a trade off solution we adopted the following procedure: pointing of the target, acquisition of a calibration lamp, putting the target in slit, acquisition of three science spectra, acquisition of a new calibration lamp, putting the target in slit, acquisition of three additional science spectra and, finally, acquisition of the last lamp.
Flat-field and bias-correction, sky-subtraction and spectrum extraction (over the innermost $\simeq 10.0\arcsec$, in the spatial direction), as well as wavelength calibration and any subsequent step of the analysis was performed with IRAF. The most relevant portion of the stacked spectrum obtained from all the 
individual usable spectra (one of the six being corrupted) is shown in Fig.~\ref{vedis}.

Unfortunately we found out, a posteriori, that significant residual velocity shifts (up to $\sim 30$~km/s) were present between the various science spectra and in the measures of the velocity difference from emission and absorption lines. Therefore, we concluded that the quality of the observational material and, in particular, the reliability of the wavelength calibration were not sufficient to reach our scientific goal, that was  challenging, given the observational set-up. 

\end{document}